# The Normalization of Co-authorship Networks in the Bibliometric Evaluation: The Government Stimulation Programs of China and Korea

**Han Woo Park *[1]  Jungwon Yoon[2]  and Loet Leydesdorff [3]**

**Abstract**

Using co-authored publications between China and Korea in *Web of Science* (WoS) during the one-year period of 2014, we evaluate the government stimulation program for collaboration between China and Korea. In particular, we apply dual approaches, full integer vs. fractional counting, to collaborative publications in order to better examine both the patterns and contents of Sino-Korean collaboration networks in terms of individual countries and institutions. We first conduct a semi-automatic network analysis of Sino-Korean publications based on the full-integer counting method, and then compare our categorization with contextual rankings using the fractional technique; routines for fractional counting of WoS data are made available at http://www.leydesdorff.net/software/fraction . Increasing international collaboration leads paradoxically to lower numbers of publications and citations using fractional counting for *performance* measurement. However, integer counting is not an appropriate measure for the evaluation of the stimulation of *collaborations*. Both integer and fractional analytics can be used to identify important countries and institutions, but with other research questions.

---

[1] Department of Media & Communication, Interdisciplinary Program of East Asian Cultural Studies, Interdisciplinary Program of Digital Convergence Business Yeungnam University, 214-1 Dae-dong, Gyeongsan-si, Gyeongsangbuk-do 712-749, South Korea; email: hanark@ynu.ac.kr *Corresponding author;

[2] National Research Foundation of Korea, 63-389 Hangang-ro 3ga, Yongsan-Gu, Seoul 140-763, South Korea; email: wonnie78@gmail.com

[3] University of Amsterdam, Amsterdam School of Communication Research (ASCoR), PO Box 15793, 1001 NG Amsterdam, The Netherlands; email: loet@leydesdorff.net

## Introduction

Today publication and innovation activities produce enormous quantities and various kinds of research data such as papers (Mehmood et al., 2016) and patents (Yoon & Park, 2016). Scientometricians have been keen to examine collaboration networks among researchers, institutions, and nation-states (Moed, 2000). One can consider co-authorships as codified markers of collaboration. Strong demand to develop a new evidence-based method for evaluation of the R&D performance of universities can be another driving factor to spread co-authorship analysis (Gautam et al., 2014). In a recent article on collaborative cultures, Kim and Park (2015, p.236) argue that co-created artifacts (e.g., co-authored articles) are crucial for facilitating cooperation at the work floor. From the development perspective of the sciences as networked communication systems, collaboration begins with shared goals (Leydesdorff, 2015). Joint writing and experimenting to claim new knowledge tends to lead to journal co-authorship in order to gain recognition via peer review and quality control. In other words, co-authorships indicate ongoing collaborative relations among academic actors engaged in a symbolic game of competition and cooperation.

Although it is hard to generalize about identifying valid data sources and reliable standard indicators for examining scholarly communication behaviors, some analytical guidelines stand out. Despite the commercial nature of the databases, *Web of Science* has been the most formal data source and a massive storehouse for publication activities including co-authorship data (Choi et al., 2015ab; Kwon et al., 2012; Leydesdorff et al., 2014; Park &

Leydesdorff, 2010). *Scopus* and *Google Scholar* are also frequently used as data sources for developing indicators. *Scopus* covers a larger set of journals including 'online first' articles from its mother company *Elsevier*, and *Google Scholar* includes non-English academic materials in various publication formats (e.g., theses, working papers, conference proceedings, book chapters, etc.) and technical formats (e.g., PDF, slide, etc) (Delgado & Repiso, 2013; Zitt, 2006) . Other specialized options for collecting publication data in specific fields include PubMed for bio-medical research, Chemical Abstracts, etc. On the other hand, it must be noted that *Web of Science* contains only a disciplinary classification at the journal level in terms of its WoS subject categories. More recently, 'altmetrics' (Bornmann, 2014; Holmberg, 2015) has emerged for citation tracking as research publications become increasingly connected via social media (Gruzd et al., 2012; Van Noorden, 2014).

An argument in support of using commercial database as a pipeline is that the inclusion criteria for journals offer an additional round of quality control (Velez-Cuartas et al., 2016) in addition to the round of quality control in the editorial process of the journal itself. Within this domain, one can further classify papers and journals in terms of their citations rates. Standardized indicators for citation have been developed. While *Web of Science* is proud of its famous indicators (e.g., ISI journal impact factor and Eigen factor score), *Scopus* has SCImago journal rank (SJR) and Source Normalized Impact per Paper (SNIP).

In a similar vein, several studies tried to standardize the measurement of the practices and trends of co-authorships. For example, King (2011), Leydesdorff and his colleagues (Leydesdorff et al., 2013, 2014; Wagner et al., 2015), and Mosbah-Natanson and Gingras (2013) conducted science mapping and data visualization to illustrate global co-authorship networks. Lemarchand (2010) also studied the scientific networks among some 12 countries where Spanish or Portuguese are predominant languages using co-authorship data. Going beyond a country-level description, Choi et al. (2015a) focused on the organization and sector levels of co-authorship networks between members of the Organisation for Economic Cooperation and Development (OECD). In Choi et al. (2015b), they have expanded their scope to university-industry-government co-publications from around 130 countries in order to examine global scholarly divide. On the other hand, Park and his colleagues (Kwon et al., 2012; Park & Leydesdorff, 2010, 2013; Shapiro et al., 2010; Shapiro & Park, 2012; So et al., 2015; Yang et al., 2010) and Shushan (2012) narrowed down their choices to single country, i.e., Korea and Singapore respectively, in terms of co-authorship over time. Likewise, Zheng et al., (2012) examined the positive impact of internationally co-authored publications on the citation performance of Chinese papers. Further, there have been some interesting approaches to discover hidden knowledge structure with a particular focus on collaboration practices within specific fields including bioinformatics (Song et al., 2013) and e-government (Khan et al., 2011) and ego-network of individual researchers' co-authorship relationship (Abbasi et al., 2012).

Given that a quality indicator for analyzing co-authorships can play a guiding role informing the research community, the choice of an adequate methodology becomes increasingly important in research management and science policies. We show in this paper that some common choices in data analysis eventually fail to capture the collaborative networks of researchers. We focus on collaboration between China and South Korea (hereafter Korea) where a number of international institutions around the world participate in joint research activities (Sun & Jiang, 2014; UNESCO, 2015).

**The Network of Sino-Korea collaborations**

The establishment of a Free Trade Agreement (FTA) between China and Korea in 2015 has opened a new era for cooperation and competition in the future. In addition to bilateral cultural, economic, and political agendas, Korea adopted China as an official partner of science and technology research. Both countries expect to raise the national competitiveness because of growth of R&D budgets and publication performance. According to UNESCO (2015), China could become the world's largest scientific publisher by 2016 and Korean publications have nearly doubled since 2005, overtaking the position of similarly populated countries like Spain.

China is the third collaborator of Korea, following after the USA and Japan, and followed by India and Germany (UNESCO, 2015). China and Korea have common interests and issues in various areas of scientific cooperation (Sun & Jiang, 2014). For example, R&D globalization and efficiency have remained unsatisfactory. Recently both countries have increased R&D investment with the objective of internationalization of domestic journals in order to gain a wider acknowledgement around the world. The Chinese government implemented a policy called the Citation Impact Upgrading Plan (CIUP) to raise the Journal Impact Factor (JIF) values of Chinese journals included in the *Web of Science* (Zhou, 2015). In a similar vein, the strong promotion policy of the Korean government induced an expanded coverage of Korean journals in *Web of Science* (Tanksalvala, 2014). The Korea Research Foundation also has a *Scopus* journal evaluation committee (KRF, 2014).

Beyond this publication policy, both China and Korea aim to achieve high-quality R&D standards because only such policies return high-tech products that can boost the national economy. The level of basic and applied scientific and technological research achievements is increasingly recognized as a primary power to move a nation from the 'catch-up' to the 'first-mover' tier (Lee, 2014). Another important complementary aspect to the Sino-Korean relationship lies in addressing global issues such as energy crises, environmental pollution, global warming, and infectious diseases.

**Two Analytical Techniques under investigation**

The network of coauthorship relations can be studied with techniques of social network analysis. A considerable number of computer programs for the analysis and visualization of networks are nowadays available (e.g., UCInet, Pajek, ORA, VOSviewer, Gephi, etc.). The mathematics underlying social network analysis is graph theory. Graphs are mainly studied as sets of nodes (vertices) and links (arc or edges). One first studies the properties of networks without considering the value of the links and then in a next step one turns to values and signs as a further extension of the proofs and algorithms. Binary networks therefore are the default in SNA. In the Drawing panel of Pajek, for example, "Forget" is the default option for "values of lines". Alternatively, one can choose for using the values as indicators of proximity or distance.

In the case of bibliometric networks the values of lines are important. One is not only interested in the collaboration between China and Korea itself, but in the intensity of the collaboration, compared, for example, with the collaboration of these two countries with the USA or other countries. The purpose of a study is often to produce a ranking. Ranking presumes that values are central. Graph-analytic measures such as centrality, however, can be very different for valued or binary networks (Brandes, 2008).

Since the early development of bibliometric indicators, furthermore, a debate has raged whether one should count publications and relations among publications with a value of one for each of them or proportionally to the number of authors, c.q. institutional addresses, involved. Mathematically, the latter way of so-called "fractional" counting has the advantage that numbers always add up to 100% (Andersen et al., 1988; Waltman & Van Eck, 2015). This may improve the consistency of indicators. Conceptually, however, one can argue that a coauthored publication can be counted as an achievement on both sides, and should thus be honored with a full point ("integer counting"). A disadvantage of fractional counting is that the numbers decline with increased collaboration, *ceteris paribus* (Leydesdorff, 1989). However, one can solve the problem that the numbers may not always add up to 100% by using relative frequencies.

To go one step further, Moed (2000) suggests that fractional counting should consider the ordinal positions of authors. In his study, the interviewed scientists are favorable of assigning higher weights to the first author because the order of co-authors reflects different proportions in the contributions, This issue becomes complicated when co-authors and their affiliation institutes have conflicting interests, for example, the recognition of the best scientists (universities) in highly competitive market for funding resources. In order to address this prolbem, Aziz & Rozing (2013) have recently introduced a measure called the 'profit-' or '$p$-index' which prioritizes the relative contribution of multiple co-authors to their publication.

In the case of a stimulation program for international collaboration such as the one here under study between China and Korea, integer counting is the obvious way to measure the success of the program; using fractional counting, international collaboration can be considered as a zero-sum game because each publication remains one full point independently of the composition of the team, whereas the objective of the program is to internationalize the team. But how would a choice for integer or fractional counting work out for the network parameters? When the networks are first considered as binary, the counting would not make a difference because the relation either exists or not. As valued networks, however, the values matter, and the way of counting may thus affect the structural parameters of networks such as centrality measures or density.

**Research questions**

We have two research questions: one substantially about the structural characteristics of the networked collaboration between China and Korea, and the remainder of the world, and secondly, about the measurement and its effects on possible conclusions.

What are the structural characteristics of international networks around the Sino-Korean collaborations? How and to what extent are full-integer or fractionally counted networks different?

What are the structural characteristics of institutional networks around the Sino-Korean collaborations? How and to what extent are full integer and fractional counting networks different?

## Method: Data Collection & Analytical Techniques

**Data**

Scientific publication data were collected from the Science Citation Index Expanded of Thomson Reuters *Web of Science* on July 10, 2015. Korea-China collaboration papers are defined as publications with at least one address in both Korea and China. The number of co-authored papers between the two countries can be identified using search queries such as "CU=(Korea AND China) AND PY=2014".[4] The retrieval includes bio-medicine as well as science & technology; but we did not include the Art & Humanities or the Social Science Citation Indices.

---

[4] The search string "CU = Korea AND PY= 2014" retrieves 63,833 records, of which 63,806 (>99.9%) has an address in South Korea and 28 in North Korea. Since this adds up to 63.834, obviously one paper was co-authored by North and South Koreans. However, one can also search with "CU= South Korea" in the database. The search "CU = (South Korea AND China) AND PY= 2014" retrieved 2,765 records on January 19, 2016.

**Methods of Integer Versus Fractional Counting Techniques**

Ever since the origins of evaluative bibliometrics, an issue has been whether a coauthored publication should be attributed as a full publication to each of the authors or rather proportionally (Narin, 1976, pp. 125f.; Small *et al*., 1985, p. 391). In the case of three authors, for example, should each of them be attributed 1/3 point or the whole number of one? Should citations then also be attributed fractionally? (Egghe, 2008; Galam, 2012). The SNIP indicator for journal evaluation (of Scopus), for example, attributes citations fractionally to journals (Moed, 2010) in order to correct for the different citation densities in fields of science (Garfield, 1979). However, this "source normalization" is from the citing side, while our focus is here on performance measurement at the cited side.

Should one also attribute publications proportionally to countries and universities? (Leydesdorff & Shin, 2011)? The issue is further complicated because the number of institutes involved can be different from the number of authors because authors may share institutional addresses. In the example above of three authors, two may come from the same institute and one from a different one: should each institute (or country) than obtain half of the credit? Or the one two-third and the other one third? The institutes can be in different countries and the question can thus be posed at all levels of aggregation.

In a debate about "the decline of British science" initiated by Irvine *et al*. (1985; Irvine & Martin, 1986), Leydesdorff (1988) argued that this "decline" was an artifact of measuring publications fractionally (cf. Anderson *et al*., 1988; Martin, 1991). Increasing collaboration at the international level leads to a decline in performance counting at the national level, *ceteris paribus*. With increasing collaboration whole numbers become fractions. Collaboration would thus be negatively incentivized (Braun *et al*., 1989). Whole number counting, however, leads to double or multiplicative counting in the case of multiple coauthorships, and then to potential inconsistencies in the evaluation. Fractional counting therefore is widely accepted among evaluators as the most appropriate normalization, because the sum-total of the citation matrix then conveniently remains 100%. For example, Waltman & van Eck (2015, at p. 892) argue that "a disadvantage of multiplicative counting is that publications do not all have the same weight in the calculation of field-normalized indicators."

In this study—occasioned by the stimulation program for Chinese-Korean collaboration recently agreed between the two governments—we propose to consider fractional and integer counting as not only two different counting schemes, but as relevant for two different systems of reference. For the reasons specified above, collaboration would be counterproductively incentivized when the efforts were evaluated using fractional counting: each of the two collaborating nations would suffer from such a scheme. Thus, the *performance* of participating agents should be accounted on the basis of integer counting. However, we shall show that integer counting is not an appropriate measure for the evaluation of the *collaboration*. In our case, a third party (e.g., the USA or Japan) may be involved, and quantitatively the links with this third country may outnumber the Sino-Korean collaboration if not weighted. Unlike the evaluation of performance, the evaluation of the collaboration requires fractional counting given our research question. In other words, the links of the networks (co-authorship relations) develop with a dynamic different from the development of the agents at the nodes. Evaluation schemes have to take these two aspects into account. The issue is not a strictly technical, but a conceptual one: the systems of reference for the evaluation are different, namely, a set of nodes or a set of links.

**Different counting rules**

The co-authorship relations add up to a network of relations. In network analysis, however, the counting rules are different. For example, in the case of a paper with three authors from institute (or country) A and two authors from institute B in another (or the same) country, the number of affiliations between the two institutes is 3 * 2 = 6. The network can be represented as a symmetrical matrix, for example, with agents (authors, institutes, or countries) on both axes (Table 1). The cell values represent the numbers of links (arcs). The single paper with three and two authors, respectively, then adds six points to the cell cross-tabling countries A and B. The symmetrical (1-mode) co-occurrence matrix can be obtained by multiplying the asymmetrical (2-mode) occurrence matrix with its transposed.

**Table 1**: Co-authorship relations using the different counting methods

|  | A | B | C |
|---|---|---|---|
| Authors | 3 | 2 | 4 |
| Institutes | a | b | c |
| Countries | * | ** | * |

**Table 1A:** Data matrix of an example of co-authorship relations in a single document

|  | A | B | C |
|---|---|---|---|
| A |  | 6 | 12 |
| B | 6 |  | 8 |
| C | 12 | 8 |  |

**Table 1B:** Affiliations matrix in SNA based on the data-matrix in Table 1A

|  | A | B | C |
|---|---|---|---|
| **Authors** | 1/3 | 2/9 | 4/9 |
| **Relations** | 9/26 | 7/26 | 10/26 |
| **Institutes** | 1/3 | 1/3 | 1/3 |
| **Countries** | ½ + ½ |  |  |

|  | A | B | C |
|---|---|---|---|
| **Authors** | 0.33 | 0.22 | 0.44 |
| **Relations** | 0.35 | 0.27 | 0.38 |
| **Institutes** | 0.33 | 0.33 | 0.33 |
| **Countries** | 0.5 + 0.5 |  |  |

**Table 1C: Different schemes for fractional counting and different levels of aggregation.**

In Table 1, we added four more co-authors from institute C in country A so that there are nine authors, three institutes, and two countries involved. Using the fractionation rule at the author level each author would obtain $1/(3 + 2 + 4) = 1/9^{th}$ point credit, divided as 3/9 for institute A, 2/9 for institute B, and 4/9 for C. The division over the countries could be 7/9 and 2/9; but one can also argue in terms of institutional addresses and then divide $2/3^{rd}$ to the one nation (A) and $1/3^{rd}$ to the other (B) or, thirdly, credit each of the two countries with 1/2. At the institutional level, each institution would then obtain $1/3^{rd}$ instead of dividing according to 3:2:4.

Note that these various options are all available in the case of each single paper. Searching in a database, however, one will always retrieve this paper as one. Network analysts are first interested in this graph in which the links among authors/institutions/countries exist or do not exist. This matrix is binary or unweighted (i.e., not valued other than with zeros and ones). In bibliometrics, the matrices are valued or, in other words, the cells are weighted in terms of the lower-triangle (or equivalently, the upper-triangle) values. The sum value of the triangles (in Table 1b) is $6 + 8 + 12 = 26$. The relative frequency of the cell {A,B}= 6/26. Since these six links are arcs in both directions (given that there are three authors from A and two from B), one can also argue for using all these values divided by 2, i.e. as edges. Since this applies to all cells, this transformation of the network does not make a difference in the computation of structural measures.

Searching samples in a database—for example, with "country = A OR country = B"—one does not retrieve the co-occurrence value between the vectors based on multiplying the mutual occurrences at the document level, but the "minimal overlap" (Morris, 2005, p.

22). For example, if the sample contains three documents with an address in A and two documents with an address in B, one retrieves a minimal overlap value of 2, and not 3 * 2 = 6 co-occurrences. According to Morris (2005, at p. 36), a representation based on the co-occurrence values is often less meaningful (for example, in co-word maps) than the one based on the minimum overlap (Zhou & Leydesdorff, in press).

In this study, we focus on (*i*) the binary matrix, (*ii*) the valued matrix which is integer counted), and (*iii*) the fractionally counted matrix, using all publications co-authored between China and Korea in 2014 as our data. We developed software for fractional counting of document sets retrieved from WoS at the author level, institutional, and national level that can be found at http://www.leydesdorff.net/software/fraction/index.htm .

## Results

**International Network**

Let us first analyze the binarized data matrices which are the basis for the overall picture of the Sino-Korean collaboration network (Table 2 & Figure 1). This set of cohesion measures was computed for both integer and fractional data using the routine in UCInet. Cohesive measures between integer and fractional networks are the same in the UCInet because the two networks are the same in terms of binary graph structure.

Various social network analysis (SNA) indicators such as network density, centralities, and geodesic distance were employed. Density value 0.667 indicates that some 66.7 percent of all possible collaborations occurred. Thus, Sino-Korean collaboration network appears to be tightly connected. Avg Degree value 82.704 reveals that about 83 out of 125 countries collaborated with each other (Freeman, 1979).

Table 2. Multiple cohesion measures of binarzed countings: International networks

| No | Metrics | Definitions | Values |
|---|---|---|---|
| 1 | Density | Number of relations divided by the maximum number of possible relations | 0.667 |
| 2 | Avg Degree | Average value of degree centralities | 82.704 |
| 3 | H-Index | Largest number x such that there are x vertices of degree at least x in the underlying graph | 80 |
| 4 | Compactness | Mean of all the reciprocal distances | 0.833 |
| 5 | Closure | Number of non-vacuous transitive triples divided by number of paths of length 2 | 0.832 |
| 6 | Avg Distance | Average geodesic distance amongst reachable pairs | 1.333 |
| 7 | SD Distance | Standard deviation of the geodesic distances amongst reachable pairs | 0.471 |
| 8 | Wiener Index | Average shortest path distance | 20662 |
| 9 | Diameter | Length of the longest geodesic distance | 2 |
| 10 | Deg Centralization | Sum of the squares of the proportion of the total centrality held by each node | 0.338 |
| 11 | Nulls | Number of cells with null values | 0.333 |
| 12 | Dependency Sum | Sum of the betweenness proportions of Y for all pairs which involve node X | 5162 |

Note: Definitions compiled and modified by the authors for this study based on various sources including Hanneman & Riddle, (2005), van Liere (2004), and http://www.analytictech.com/ucinet/help/idx.htm

*Calculated using "Multiple Cohesion Measures" option in the UCInet 6 Version 6.590

Figure 1. Sino-Korea collaboration international networks: Binary matrix

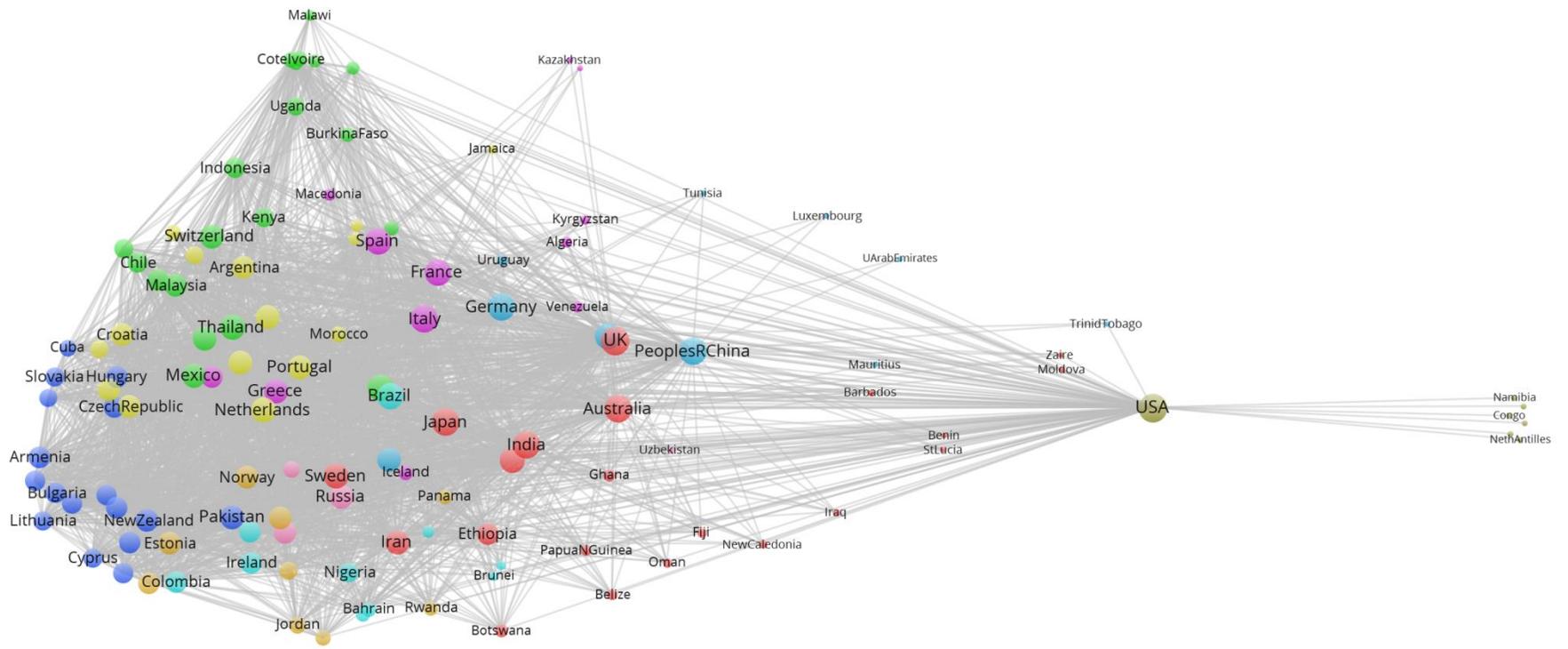

Table 2. Multiple cohesion measures of binarzed countings: International networks

Figure 1. Sino-Korea collaboration international networks: Binary matrix

In many ways, a fractionally counted network is very different from an integer graph. Table 3 compares multiple structural measures. The table contains a list of metrics related to the degree and normalized degree (NrmDegree) centralities, together with the share (expressed as a percentage), for each network. While ten metrics decrease in degree centralities, only two measures increase. Noticeably, both 'Blau Heterogeneity' and 'Normalized(IQV)' values increased. Because we considered the relative portion of collaboration in the fractional method, these differences occurred. This also generated a contrasting network structure as visualized in Figure 2 and 3.

Table 3. A comparison between integer and fractional countings

The Sino-Korean collaboration network contains 125 countries and 255 relations in terms of dichotomy. The degree centralities of individual countries vary widely, as summarized in Table 4. In an integer counted network, the top 10 countries include the USA, Italy, Turkey, Russia, India, Germany, France, and the UK. Korea and China are ranked 9th and 11$^{th}$, respectively. This would indicate that neither country plays the most productive roles in the ego-networks of their collaboration ties.

Table 4. Normalized degree centralities between two international networks in 2014

Table 3. A comparison between integer and fractional countings

| No | Metrics | Integer | | | Fractional | | | Difference | | | Binary | | |
|---|---|---|---|---|---|---|---|---|---|---|---|---|---|
| | | Degree | NrmDegree | Share | Degree | NrmDegree | Share | Degree | NrmDegree | Share | Degree | NrmDegree | Share |
| 1 | Mean | 41284.543 | 0.181 | 0.008 | 21.920 | 0.026 | 0.008 | -99.947% | -85.635% | 0.000% | 82.704 | 66.697 | 0.008 |
| 2 | Std Dev | 115056.391 | 0.504 | 0.022 | 113.768 | 0.136 | 0.042 | -99.901% | -73.016% | 90.909% | 29.410 | 23.718 | 0.003 |
| 3 | Sum | 5160568.000 | 22.623 | 1.000 | 2740.029 | 3.273 | 1.000 | -99.947% | -85.532% | 0.000% | 10338.000 | 8337.097 | 1.000 |
| 4 | Variance | 13237971968.000 | 0.254 | 0.000 | 12943.205 | 0.018 | 0.002 | -100.000% | -92.913% | 0.200% | 864.960 | 562.539 | 0.000 |
| 5 | SSQ | 1867798282240.000 | 35.894 | 0.070 | 1677962.625 | 2.394 | 0.223 | -100.000% | -93.330% | 218.571% | 963114.000 | 626374.875 | 0.009 |
| 6 | MCSSQ | 1654746513408.000 | 31.800 | 0.062 | 1617900.625 | 2.308 | 0.215 | -100.000% | -92.742% | 246.774% | 108120.047 | 70317.406 | 0.001 |
| 7 | Euc Norm | 1366674.125 | 5.991 | 0.265 | 1295.362 | 1.547 | 0.473 | -99.905% | -74.178% | 78.491% | 981.384 | 791.438 | 0.095 |
| 8 | Minimum | 3.000 | 0.000 | 0.000 | 0.007 | 0.000 | 0.000 | -99.767% | 0.000% | 0.000% | 2.000 | 1.613 | 0.000 |
| 9 | Maximum | 969793.000 | 4.251 | 0.188 | 892.244 | 1.066 | 0.326 | -99.908% | -74.924% | 73.404% | 124.000 | 100.000 | 0.012 |
| 10 | N of Obs | 125.000 | 125.000 | 125.000 | 125.000 | 125.000 | 125.000 | 0.000% | 0.000% | 0.000% | 125.000 | 125.000 | 125.000 |
| 11 | Network Centralization | 4.140% | | | 1.060% | | | -74.396% | N.A. | N.A. | 33.840% | | |
| 12 | Blau Heterogeneity | 7.010% | | | 22.350% | | | 218.830% | N.A. | N.A. | 0.900%. | | |
| 13 | Normalized (IQV) | 6.260% | | | 21.720% | | | 246.965% | N.A. | N.A. | 0.100% | | |

*Calculated using Degree Centrality (old) option in the UCInet 6 Version 6.590

Table 3 - Appendix. Definitions of metrics measuring structural properties of entire network

| No | Metrics | Definitions |
|----|---------|-------------|
| 1 | Degree | Number of immediate ties that a node has, rather than indirect ties to all others in the network |
| 2 | Normalized degree (NrmDegree) | Degree divided by (n-1) times a 100. It is a percentage of centrality that you can maximally have. It is used to compare centrality between two different network seizes |
| 3 | Share | Centrality measure of the node divided by the sum of all the node centralities in the network |
| 4 | Mean | Average value |
| 5 | Standard Deviation (Std Dev) | Amount of variation or dispersion of a set of values |
| 6 | Sum | Total amount of values |
| 7 | Variance | One could examine whether the variability is high or low relative to the typical scores by calculating the coefficient of variation (standard deviation divided by mean, times 100) for degree centrality |
| 8 | Sums of SQuares (SSQ) | Total amount of the squares of the differences from the mean, being used as part of a standard way of evaluating randomness (or co-variance) of results |
| 9 | Mean Centered Sum of Squares (MCSSQ) | Nearly similar to SSQ values in SNA, generally subtracting average from actual values in the network |
| 10 | Euclidean Norm (Euc Norm) | Length of the shortest (the most straight-line) distance between a pair of nodes in the network |
| 11 | Minimum | The lowest value |
| 12 | Maximum | The highest value |
| 13 | N of Obs | Total number of nodes |
| 14 | Network Centralization | Expressed as a percentage, centralization reveals particular properties of the network structure as a whole. Centralization refers to the overall cohesion or integration of the network. Networks may, for example, be more |

|  |  | or less centralized around particular nodes or sets of nodes. |
|---|---|---|
| 15 | Blau Heterogeneity | Sum of the squares of the proportion of the total centrality held by each node |
| 16 | Normalized Index of Qualitative Variation (IQV) | Normalized version of Heterogeneity value |

Note: Definitions compiled and modified by the authors for this study based on various sources including Hanneman & Riddle, (2005), van Liere (2004), and http://www.analytictech.com/ucinet/help/idx.htm

Table 4. Normalized degree centralities between two international networks in 2014

| Rank | Country | nDegree-integer | Country | nDegree-Fractional |
|---|---|---:|---|---:|
| 1 | U.S.A. | 7820.911 | PeoplesRChina | 7.193 |
| 2 | Italy | 5670.097 | SouthKorea | 7.111 |
| 3 | Turkey | 2055.847 | U.S.A. | 2.306 |
| 4 | Russia | 1719.274 | Japan | 0.582 |
| 5 | India | 1707.645 | Italy | 0.429 |
| 6 | Germany | 1656.548 | Germany | 0.401 |
| 7 | France | 1617.726 | UK | 0.311 |
| 8 | SouthKorea | 1512.185 | France | 0.293 |
| 9 | UK | 1302.492 | India | 0.285 |
| 10 | PeoplesRChina | 871.468 | Taiwan | 0.258 |

| 11 | Belgium | 857.645 | Russia | 0.249 |
|----|---------|---------|--------|-------|
| 12 | Brazil | 853.379 | Australia | 0.223 |
| 13 | Mexico | 843.008 | Canada | 0.199 |
| 14 | Switzerland | 836.242 | Spain | 0.149 |
| 15 | Spain | 833.169 | Singapore | 0.142 |
| 16 | Egypt | 823.895 | SaudiArabia | 0.101 |
| 17 | Hungary | 664.863 | Turkey | 0.099 |
| 18 | Finland | 587.734 | Switzerland | 0.097 |
| 19 | Iran | 581.815 | Poland | 0.085 |
| 20 | Poland | 577.419 | Brazil | 0.081 |
| 21 | Japan | 542.565 | Netherlands | 0.080 |

Figure 2. Sino-Korea collaboration international networks: Integer

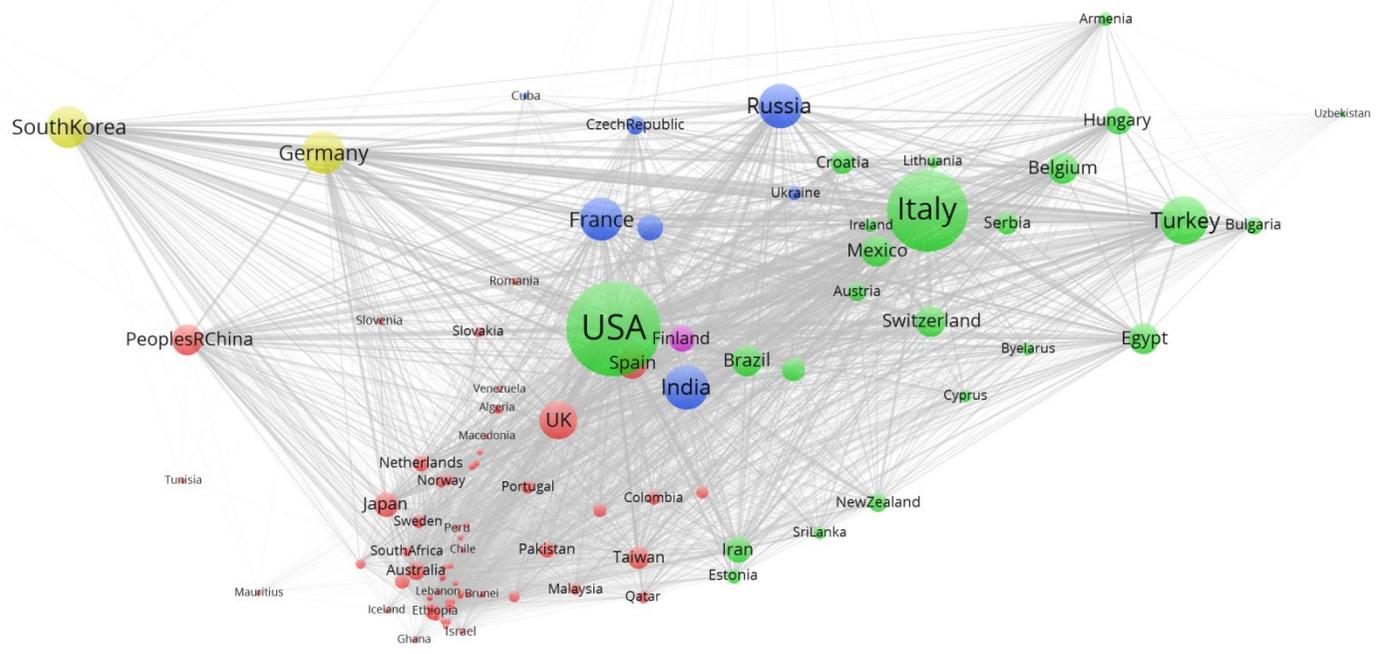

A closer examination of a fractional network, however, prevents this erroneous conclusion from being drawn. China is the most central country, followed by Korea in the network based on fractional counting (Figure 3). Interestingly enough, Asian countries (Japan, India, Taiwan, Singapore) occupy higher positions, compared to their marginal positions in the other integer network. United Arab Emirates (henceforth, UAE) had the largest occurred discrepancy from 125th in the integer network to 68th in the fractional network (+57). When analyzing the networked position of UAE in the integer network, it has relations only with China and Korea, making a closed triad structure. Trinidad and Tobago, Luxembourg, and Tunisia follow after the UAE in terms of the biggest change in their ranks: 124th to 70th (+54), 123rd to 71st (+52), and 122nd to 75th (+47), respectively. Interestingly, only three countries, UAE, Luxembourg, and Tunisia are isolated from the ego network of U.S.A. Trinidad and Tobago has connections only with China (1 tie), Korea (2 ties), and the U.S.A. (2 ties).

Figure 3. Sino-Korea collaboration international networks: Fractional

Figure 3. Sino-Korea collaboration international networks: Fractional

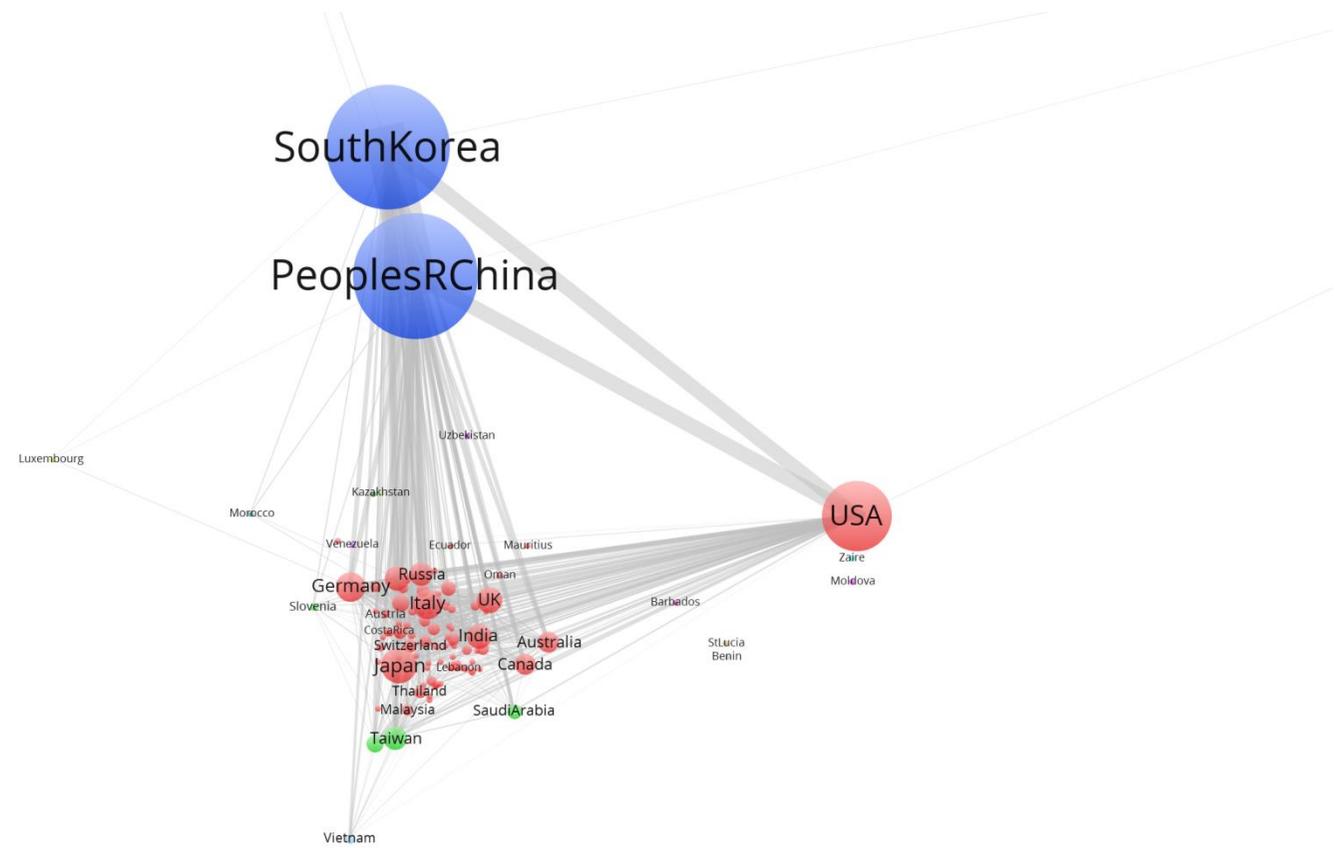
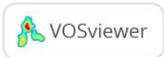

## Institutional Network

While Table 5 provides multiple cohesion measures in UCInet for the binary counted matrix in institutional networks, Table 6 summarizes the structural measures obtained using the Degree option in the UCInet menu. Nearly 100% is consistent across the corresponding metrics in Tables 3 and 6.

Table 5. Multiple cohesion measures of binarized countings: Institutional networks

Table 6. A comparison between integer and fractional countings

Table 5. Multiple cohesion measures of binarzed countings: Institutional networks

| No | Metrics | Values |
|---|---|---|
| 1 | Density | 0.031 |
| 2 | Avg Degree | 136.593 |
| 3 | H-Index | 388 |
| 4 | Compactness | 0.442 |
| 5 | Closure | 0.493 |
| 6 | Avg Distance | 2.404 |
| 7 | SD Distance | 0.6 |
| 8 | Wiener Index | 46815544 |
| 9 | Diameter | 6 |
| 10 | Deg Centralization | 0.384 |
| 11 | Nulls | 0.969 |
| 12 | Dependency Sum | 27345362 |

Table 6. A comparison between integer and fractional countings

| No | Metrics | Integer | | | Fractional | | | Difference | | | Binary | | |
|---|---|---|---|---|---|---|---|---|---|---|---|---|---|
| | | Degree | NrmDegree | Share | Degree | NrmDegree | Share | Degree | NrmDegree | Share | Degree | NrmDegree | Share |
| 1 | Mean | 1295.636 | 0.010 | 0.000 | 0.760 | 0.001 | 0.000 | -99.941% | -90.000% | N.A. | 136.593 | 3.085 | 0.000 |
| 2 | Std Dev | 5469.135 | 0.041 | 0.001 | 2.967 | 0.006 | 0.001 | -99.946% | -85.366% | 0.000% | 192.387 | 4.346 | 0.000 |
| 3 | Sum | 5737076.000 | 43.011 | 1.000 | 3365.467 | 6.412 | 1.000 | -99.941% | -85.092% | 0.000% | 604836.000 | 13662.436 | 1.000 |
| 4 | Variance | 29911438.000 | 0.002 | 0.000 | 8.802 | 0.000 | 0.000 | -100.000% | -100.000% | 0.200% | 37012.883 | 18.886 | 0.000 |
| 5 | SSQ | 139881005056.000 | 7.862 | 0.004 | 41532.477 | 0.151 | 0.004 | -100.000% | -98.079% | 0.000% | 246509712.000 | 125780.984 | 0.001 |
| 6 | MCSSQ | 132447846400.000 | 7.444 | 0.004 | 38974.578 | 0.141 | 0.003 | -100.000% | -98.106% | -25.000% | 163893056.000 | 83626.031 | 0.000 |
| 7 | Euc Norm | 374006.688 | 2.804 | 0.065 | 203.795 | 0.388 | 0.061 | -99.946% | -86.163% | -6.154% | 15700.628 | 354.656 | 0.026 |
| 8 | Minimum | 1.000 | 0.000 | 0.000 | 0.007 | 0.000 | 0.000 | -99.300% | 0.000% | 0.000% | 1.000 | 0.023 | 0.000 |
| 9 | Maximum | 246755.000 | 1.850 | 0.043 | 72.872 | 0.139 | 0.022 | -99.970% | -92.486% | -48.837% | 1834.000 | 41.428 | 0.003 |
| 10 | N of Obs | 4428.000 | 4428.000 | 4428.000 | 4428.000 | 4428.000 | 4428.000 | 0.000% | 0.000% | 0.000% | 4428.000 | 4428.000 | 4428.000 |
| 11 | Network Centralization | 1.840% | | | 0.140% | | | -92.391% | N.A. | N.A. | 38.360% | | |
| 12 | Blau Heterogeneity | 0.420% | | | 0.370% | | | -11.905% | N.A. | N.A. | 0.070% | | |
| 13 | Normalized (IQV) | 0.400% | | | 0.340% | | | -15.000% | N.A. | N.A. | 0.040% | | |

Table 7. Normalized degree centralities between two institutional networks in 2014

| Institution | nDegree-Integer | Rank | Institution | nDegree-Fractional |
|---|---|---|---|---|
| IstNazlFisNucl | 55.739 | 1 | ChineseAcadSci | 0.016 |
| RheinWestfalThAachen | 11.843 | 2 | SeoulNatlUniv | 0.016 |
| UnivBelgrade | 8.366 | 3 | HanyangUniv | 0.011 |
| InstHighEnergyPhys | 7.356 | 4 | YonseiUniv | 0.010 |
| SezioneIstNazlFisNucl | 7.330 | 5 | SungkyunkwanUniv | 0.009 |
| UnivRome | 6.764 | 6 | InhaUniv | 0.008 |
| InstTheoretExptphys | 6.383 | 7 | KoreaUniv | 0.008 |
| CERN | 6.271 | 8 | GyeongsangNatlUniv | 0.008 |
| KyungpookNatlUniv | 5.967 | 9 | ChungnamNatlUniv | 0.008 |
| UnivKansas | 5.966 | 10 | ChonbukNatlUniv | 0.008 |
| UnivPerugia | 5.924 | 11 | PukyongNatlUniv | 0.008 |
| UnivAthens | 5.881 | 12 | PusanNatlUniv | 0.007 |
| JointInstNuclRes | 5.723 | 13 | KyungHeeUniv | 0.007 |
| PurdueUniv | 5.674 | 14 | PekingUniv | 0.006 |
| Cnrs | 5.654 | 15 | KoreaAdvInstSciTechnol | 0.006 |
| RussianAcadSci | 5.553 | 16 | YanbianUniv | 0.006 |
| WayneStateUniv | 5.489 | 17 | ChungbukNatlUniv | 0.005 |
| Caltech | 5.459 | 18 | UnivHongKong | 0.005 |
| MoscowMvLomonosovStateUniv | 5.439 | 19 | ShanghaiJiaoTongUniv | 0.005 |
| PanjabUniv | 5.393 | 20 | TsinghuaUniv | 0.005 |
| KoreaUniv | 5.379 | 21 | YeungnamUniv | 0.005 |

Table 7 shows the degree centrality values in the system of Sino-Korean collaborations in the top 20 institutions out of 4428 institutional addresses mentioned in the bylines of the *Web of Science* publications. According to the integer counting analysis, CERN's performance (ranked 8th) was successful in mediating institutional collaboration. In contrast, CERN CERN occupies the 360th position in the fractional counting rank, a decrease of 352 steps. On the other hand, the Chinese Academy Science (Chinese AcadSci) becomes the new leader in collaboration, jumping from the 264th to the 1st place. Next is Seoul National University (SeoulNatlUniv) that moved from 190th to 2nd place.

Table 7. Normalized degree centralities between two institutional networks in 2014

Figure 4. Sino-Korea collaboration institutional networks: Integer

Figure 5. Sino-Korea collaboration institutional networks: Fractional

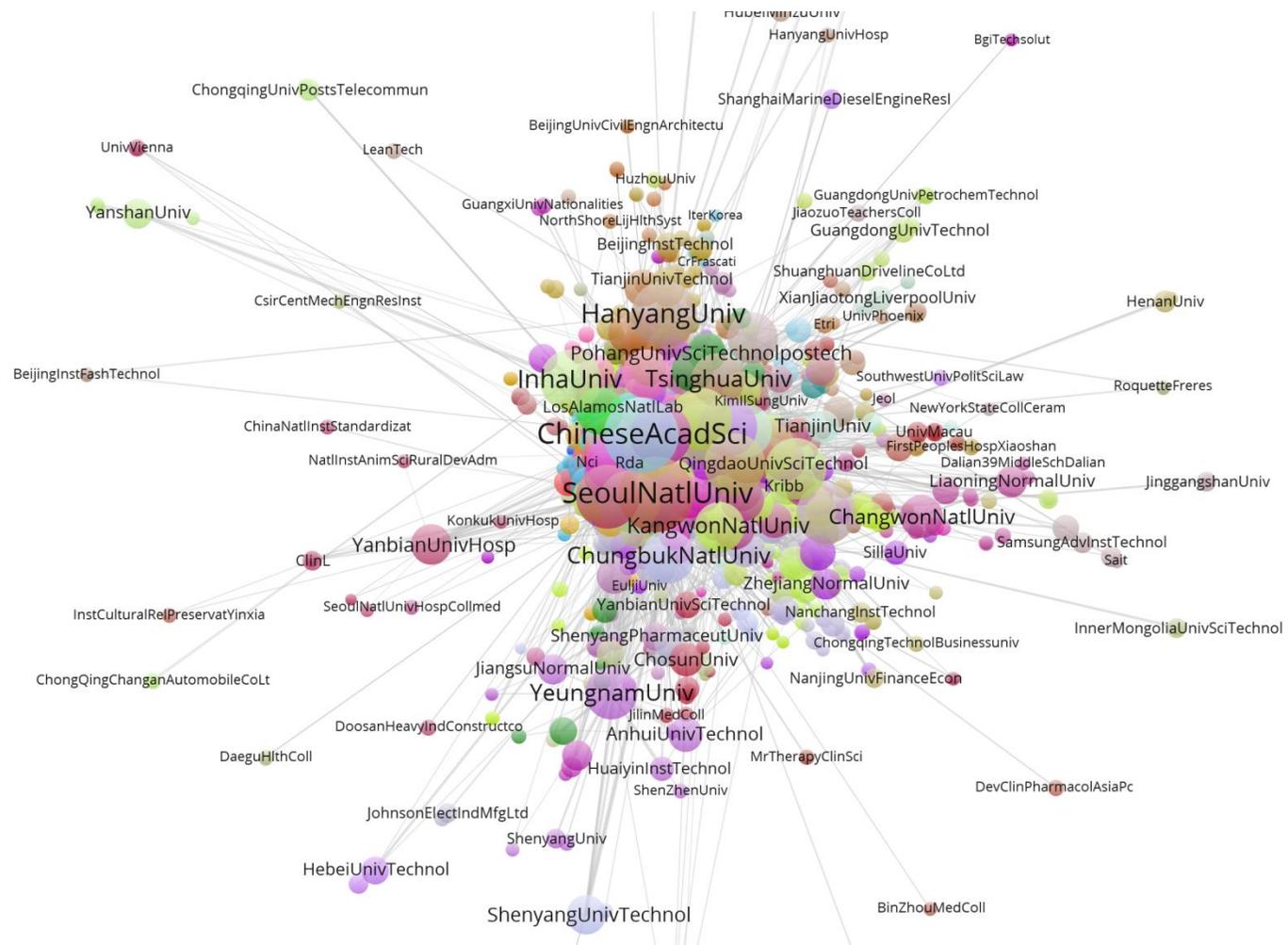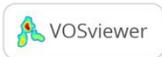

This shows that the choice of a counting method is an important factor in evaluating collaboration activity. The implication is that there are many more changes occurring in institutional networks, whereas countries in the international network showed relatively fewer movements. However, more dramatic differences occurred for CERN—as expected—but also for the Chinese Academy of Science and Seoul National University. Thus, if we neglect the normalization effect of collaboration on performance evaluation, researchers and policymakers might not paint a more complete picture of the institutional networks.

## Discussion and Conclusions

This analysis of the collaboration patterns among countries belonging to the Sino-Korea research network reveals some interesting facets of how scientometric data can be used to map international and institutional co-authorship culture. Our findings demonstrate that scholarly documents of research collaboration, in the form of co-authorships, exhibit significant characteristics that attract many other participants in Sino-Korea collaboration: the networked practices of science and affiliations among countries and institutions. One of the main challenges of collaboration mapping lies in evaluating individual contribution on the quantitative scale. Social network analysis based on graph theory has offered a useful tool to examine co-authorships represented in scientific publications. However, a comparative exercise in current scientometric provides an important implication. The results differ widely between integer counting derived from the traditional graph-theoretical network approaches and the new fractional method based on scientometrics. For example, we could have ignored a major effect of fractional counting because of the numerous authors involved at centers like CERN. Therefore, this study attempted to detail this issue for SNA-dominated co-authorship studies. A singular focus on the network graph obscures a key point. From a perspective of the measurement instrument, future studies are needed for careful comparison among the various measures.

Differences in data analysis technique may cause different research results. More importantly, it is hard to evaluate the validity of certain frequently used statistical analyses within a single study. For example, the rank-ordering comparison between integer and fractional centralities in international networks for 125 countries reveals the two rankings to be extremely similar (r = 0.935, p < .01). Furthermore, the Quadratic Assignment Procedure (QAP, Dekker et al., 2007) correlation indicates that the two networks are significantly similar in terms of their internal matrices structures with coefficients .102 (p < .01), These results were cross-checked using Pajek, another SNA software. In spite of both measures of the ego-network of Sino-Korea collaboration, as described in the Results section, such statistically significant values comparing the two networks are also misleading. As emphasized in Tables 1 and 3 (comparing multiple cohesion measures between two measurement methods) the following questions require further exploration. Which measures in SNA can be used for co-authorship and/or citation analysis and sometimes why not? What are the limitations? The approximate 40 percentage difference between integer and fractional networks is enormous, the results also make no sense in some of the cases, and the results differ widely depending on the methodology.

This research provides a primary case study that establishes a reliable methodological approach for using publication data in the globalized research system. Furthermore, it sheds light on the ways in which, at least to some degree, SNA-mediated methods serve as a science-mapping tool to organize data for co-authorship analysis, capture collaboration activities on many levels, and reflect the academic landscape of international and/or institutional cooperation and competition. Our claim, however, is not that graph-theoretic methods are suspect. QAP as a matrix correlation and regression technique has been used by network researchers for a long time. Because humans tend to see what they want to see in results, integer network analysis alone can be problematic, and even if the sophisticated network visualization has greater credibility than traditional tables and charts, the statistics needs to be complemented so the single scale of the parameters does not bias the results.

Researchers collect, classify, curate, visualize, and discuss data as evidence to evaluate prior literature and develop a new body of knowledge. However, there some 'tension' exists between a common practice and a new approach because a particular framework is widely recognized within a shared 'paradigm' (Kuhn, 1970) that is acknowledged in a particular academic community at a certain period. In this regard, Thelwall (2004) argued that the correctness of any methodological technique is socially constructed, not naturally given. In line with these arguments, it may be up to the researcher to select the appropriate indicators for the data under investigation.